\documentclass[pra,aps,twocolumn,superscriptaddress,showpacs,longbibliography]{revtex4-2}
\usepackage{graphicx}
\usepackage{amssymb}
\usepackage{wrapfig}
\usepackage{epstopdf}
\usepackage{amsmath}
\usepackage{xspace} 
\usepackage{color}
\newcommand{\cor}{\textcolor{black}}

\begin{document}
\title{Quantum coherent states of mass-imbalanced electron-hole system within optical microcavities}
\author{Thi-Hau Nguyen}
\affiliation{Faculty of Physics, Graduate University of Science and Technology, Vietnam Academy of
Science and Technology, 18 Hoang Quoc Viet, Hanoi 10072, Vietnam.}
\affiliation{Department of Physics, Hanoi University of Mining and Geology, Duc Thang, Bac Tu Liem, Hanoi 10072, Vietnam;}
\author{Thi-Hong-Hai Do}
\affiliation{Department of Physics, Hanoi University of Mining and Geology, Duc Thang, Bac Tu Liem, Hanoi 10072, Vietnam;}
\author{Van-Nham Phan}
\thanks{{Corresponding author: phanvannham@duytan.edu.vn}}
\affiliation{Institute of Research and Development, Duy Tan University, 3 Quang Trung, Danang  550000, Vietnam}
\affiliation{Faculty of Natural Sciences, Duy Tan University, 3 Quang Trung, Danang 550000, Vietnam}

\begin{abstract}
The interplay of the excitonic-like polariton, polariton and photonic-like polariton coherent states in mass-imbalanced electron-hole systems within optical microcavities is \cor{theoretically} examined. Utilizing the unrestricted Hartree-Fock approximation, we derive a set of self-consistent equations that evaluate the excitonic and photonic order parameters in a two-band electronic model, accounting equally for both electron-hole Coulomb attraction and light-matter coupling. Analyzing the competition among these condensate order parameters reveals a complex phase structure of coherent states in the ground state. As the mass imbalance is reduced, we observe a transition from a normal disordered electron-hole-photon system to excitonic-like, polariton, and ultimately photonic-like polariton condensation states. The distinct features of these robust condensates can be identified in the momentum distribution of the electron-hole pair amplitude and the photonic density, as well as in the wave-number-resolved photoemission spectra of electrons, holes, and photons. Increasing the excitation density further expands the range of condensation states. Additionally, lowering the mass imbalance leads to the emergence of quantum coherent bound states prior to the formation of robust condensates, which are evidenced by the static and dynamical excitonic and photonic susceptibility functions.
\end{abstract}

\date{\today}
\maketitle

\section{Introduction}
Bose-Einstein condensation (BEC), a coherent quantum state of an indistinguishable bosonic quasiparticle system is always one of the most attractive issues in condensed matter physics. At sufficiently low temperatures, an amount of the bosonic quasiparticles can be condensed in a single coherent quantum state and described by a single wave function. The critical temperature for the condensation state is inversely approximate to the mass of a quasiparticle. In this sense, in order to stabilize the BEC in practice one needs to establish a large bundle of sufficiently long lifetime and small effective mass bosonic quasiparticles. Exciton, a composite (a bound state) of electron and hole due to the Coulomb interaction is one of the bosonic quasiparticles with the effective mass in the order of $10^{-1}m_e$ ($m_e$ is the free electron mass)~\cite{MS00}. \cor{The effective mass} is much lower than that of the lightest neutral atoms such as hydrogen \cor{and these excitons are possibly condensed at temperatures around mK, however, these temperatures are still very small}~\cite{Im20,RMP.82.1489}. In the case that the excitons are coupled to photons in a microcavity, a strong matter-light interaction might appear and a new composite bosonic quasiparticle so-called polariton is released~\cite{Im20}. Due to its extremely small effective mass (in a range of $10^{-5}m_e$), one might expect to observe the BEC of the polaritons even at room temperature and open up the possibility of great application in modern technologies such as optoelectronics and quantum information~\cite{Im20}. Indeed, room-temperature polaritonic BEC has been observed in various two-dimensional (2D) material systems such as transition metal dichalcogenides (TMD)~\cite{YHA2019,JRA2021,Zhao2023}, perovskites~\cite{RJJ2018,RSJ2020,Jiepeng2024,Wu2024}, and organics~\cite{JTL2014,MSO2020,MWK2022} fabricated in an optical microcavity. Inspecting the polariton condensation states in a microcavity has thus stimulated much interest, recently.

In an optical microcavity, exciton induced from a quantum well is an important matter excitation in the polariton formation. Enlightening the physical properties of the excitons is thus extremely necessary in exploring the formation and stability of the polariton condensation state. As reported in the literature, excitons can be formed and condensed in various 2D materials and also in double layer systems~\cite{NC.5.4555,NatCommu.12.1969,npjQM.6.52,PRB.104.L121201}. The crucial points distinguishing the difference among the materials affecting the exciton formation are a mass imbalance of the electrons in the conduction and holes in the \cor{valence} bands and the bandgap between the bands. For a TMD monolayer as an example, a ratio between the mass of electrons ($m_e$) and the mass of holes ($m_h$) is $m_e/m_h\approx 1$~\cite{2DM.2.022001,Jiwon2014}, in the meanwhile, $m_e/m_h\approx 0.5$ for CdTe~\cite{TTDBao2006} or ZnO~\cite{JPC.100.1027}, $m_e/m_h\approx 0.15$ for a GaAs~\cite{PhysicaB.210.1,SGJ1995}, or GaN~\cite{Suzuki1995}, and even $m_e/m_h\approx 0.1$ for AlN quantum wells~\cite{Suzuki1995}. The effective mass of the \cor{valence} holes thus varies in a wide range from extremely heavy to comparable to that of the conduction electrons. Examining the impact of the mass imbalance on the formation and condensation state of excitons in semimetal and semiconducting materials has been concentrated recently~\cite{PRB.95.045101,CMP.23.43709}. Interpreting the 2D electron-hole gas in the extended Falicov-Kimball model (EFKM), one has specified the extension of the excitonic condensation state by lowering the mass imbalance~\cite{PRB.107.115106,PRB.109.085105,PRB.110.235143}. The EFKM has been proved as one of the most effective models addressing the formation and stability of the exciton in semimetal-semiconducting transition materials~\cite{BGBL04,PBF10}. 

The mass imbalance of the electrons and holes in a microcavity has been examined recently~\cite{PRB.104.245404,PRR.2.023089,PhDTh-Tiene}. These works have specified the significant roles of mass imbalance on electron-hole pair promotion and polariton scattering, but no bound coherent state is considered. In this work, we focus on specifying the formation and condensation states of polaritons and also their competition with other condensates such as excitonic and photonic condensation states in a microcavity under the impact of the mass imbalance. Without mass imbalance, i.e., $m_e=m_h$, the polariton condensation states in microcavities have been intently inspected~\cite{Kamide2011,PRL.111.026404,PRR.1.033120,NHAMB2016,NHAM2016,NHAM2017,NN2019}. However, as mentioned above, these studies are applicable only for the TMD monolayer quantum well situation, they thus could not be used to address the exciton-polariton signatures in a microcavity in general. In a lattice, the mass of a carrier is inversely proportional to its hopping term. In the unit of the hopping integral of the conduction electrons, the hopping integral of the \cor{valence} holes can be considered as the mass imbalance factor~\cite{PRB.107.115106,PRB.109.085105,PRB.110.235143}. Analyzing fluctuation signatures of the excitons, polaritons, and photons versus the hopping integral of the \cor{valence} holes would deliver us the impact of the mass imbalance on the condensation states in a microcavity. The electron-hole-photon system in the microcavity is described by a model in which the electrons and holes are addressed in the two-band energy Hamiltonian with the local interband Coulomb interaction. The matter-light coupling is treated on an equal footing with the Coulomb interaction. In this sense, the excitonic-photonic resonance would be fully considered to specify the formation and condensation of polaritons, and also their competition to that of excitons and photons in microcavities. The many-particle Hamiltonian here is not exactly solvable and we need to use some approximations. In this work, the Hamiltonian is analyzed in the framework of the unrestricted Hartree-Fock approximation (UHFA). As a generation of the original Hartree-Fock approximation, the UHFA takes into account decoupling with respect to all off-diagonal expectation values, a complex of the spontaneous symmetry breaking in the system described by the Hamiltonian would be delivered~\cite{SC08}. The UHFA has been widely used to investigate the phase structures of the excitonic condensation states in semimetal-semiconducting transition materials, also for the mass imbalance situation~\cite{PRB.107.115106,PRB.109.085105,PRB.110.235143}, or the mass equivalent exciton-polariton condensates in microcavities~\cite{NHAM2016,NHAM2017,NN2019}. In this work, the mass imbalance influence on the formation and stability of the condensation states of polaritons in microcavity in the ground state is inspected. With the help of the random phase approximation (RPA), both the excitonic and photonic susceptibilities of the mass imbalance electron-hole-phonon system would be evaluated. The polaritonic bound state fluctuations then could be discussed in the signatures of the low-energy excitations. Our results thus would release a significant picture addressing the nature of the competition of the excitonic-like polariton, photonic-like polarition, and polariton condensation states and also their tendencies when the system is out of the order states.

This paper is organized as follows. In Sec. II, we address the equilibrium mass imbalance electron-hole-photon system in a microcavity and its self-consistent equations finding condensate order parameters in the framework of the unrestricted Hartree-Fock approximation. The analytical solutions of the excitonic and photonic dynamical susceptibility functions are given in the random phase approximation. Numerical results of the condensate order parameters and the static and dynamical susceptibilities are presented in Sec. III. The phase diagrams of the excitonic-like polariton, polariton, and photonic-like polariton condensates and the bound state fluctuations in the ground state in the impact of the mass imbalance are discussed. The final section summarizes and concludes the work.

\section{Hamiltonian and theoretical approaches}

\subsection{Hamiltonian}
To investigate the polariton condensation states in a microcavity, we assume that the system is in thermal equilibrium. In general, the polaritons are not isolated from their surrounding and the exciton-polariton system cannot establish the thermal dynamic equilibrium state. However, due to improving the quality of the microcavity fabrication, the polariton lifetime might be enhanced and it would be larger than thermalization time, especially in the ground state~\cite{Im20}. In this situation, the system can be idealized in the thermal equilibrium~\cite{SUN2017}. The chemical potential can be established and the condensates are able to be settled in a long range over the whole system. In the momentum space, the Hamiltonian describing the electrons, holes, and photons composited in a microcavity thus can be written as follows:
\begin{equation}\label{eq1}
\begin{split}
\mathcal{H} &= \sum_{\mathbf{k}} \varepsilon_{\mathbf{k}}^e e_{\mathbf{k}}^\dagger e^{}_{\mathbf{k}} + \sum_{\mathbf{k}} \epsilon_{\mathbf{k}}^h h_{\mathbf{k}}^\dagger h^{}_{\mathbf{k}} + \sum_{\mathbf{q}} \omega^{}_{\mathbf{q}} \psi_{\mathbf{q}}^\dagger \psi^{}_{\mathbf{q}}\\
 &\quad - \frac{U}{N} \sum_{\mathbf{k}_1\mathbf{k}_2\mathbf{q}} e_{\mathbf{k}_1+\mathbf{q}}^\dagger e^{}_{\mathbf{k}_1} h_{\mathbf{k}_2-\mathbf{q}}^\dagger h^{}_{\mathbf{k}_2}\\
 &\quad - \frac{g}{\sqrt{N}} \sum_{\mathbf{k}\mathbf{q}} (e_{\mathbf{k}+\mathbf{q}}^\dagger h_{-\mathbf{k}}^\dagger \psi^{}_{\mathbf{q}} + \textrm{H.c.}),
\end{split} 
\end{equation}
where $e_{\mathbf{k}}^\dagger (e_{\mathbf{k}})$, $h_{\mathbf{k}}^\dagger (h_{\mathbf{k}})$ and $\psi_{\mathbf{q}}^\dagger (\psi_{\mathbf{k}})$ respectively are the creation (annihilation) operators of the spinless electrons in the conduction band, holes in the valence band and photons with momentum ${\mathbf{k}}$. The first three terms thus describe a non-interacting electron-hole-photon system in a microcavity, with the dispersions of the electrons (holes)
\begin{equation}\label{eq2}
\varepsilon_{\mathbf{k}}^{e(h)} = - t^{e(h)}\gamma_{\mathbf{k}}+\frac{E_g + 8t^{e(h)} - \mu}{2},
\end{equation}
and photons
\begin{equation}\label{eq5}
\omega_{\mathbf{q}} = \sqrt{\mathbf{q}^2 + \omega_c^2} - \mu.
\end{equation}

\begin{figure}[t]
\includegraphics[width=0.450\textwidth]{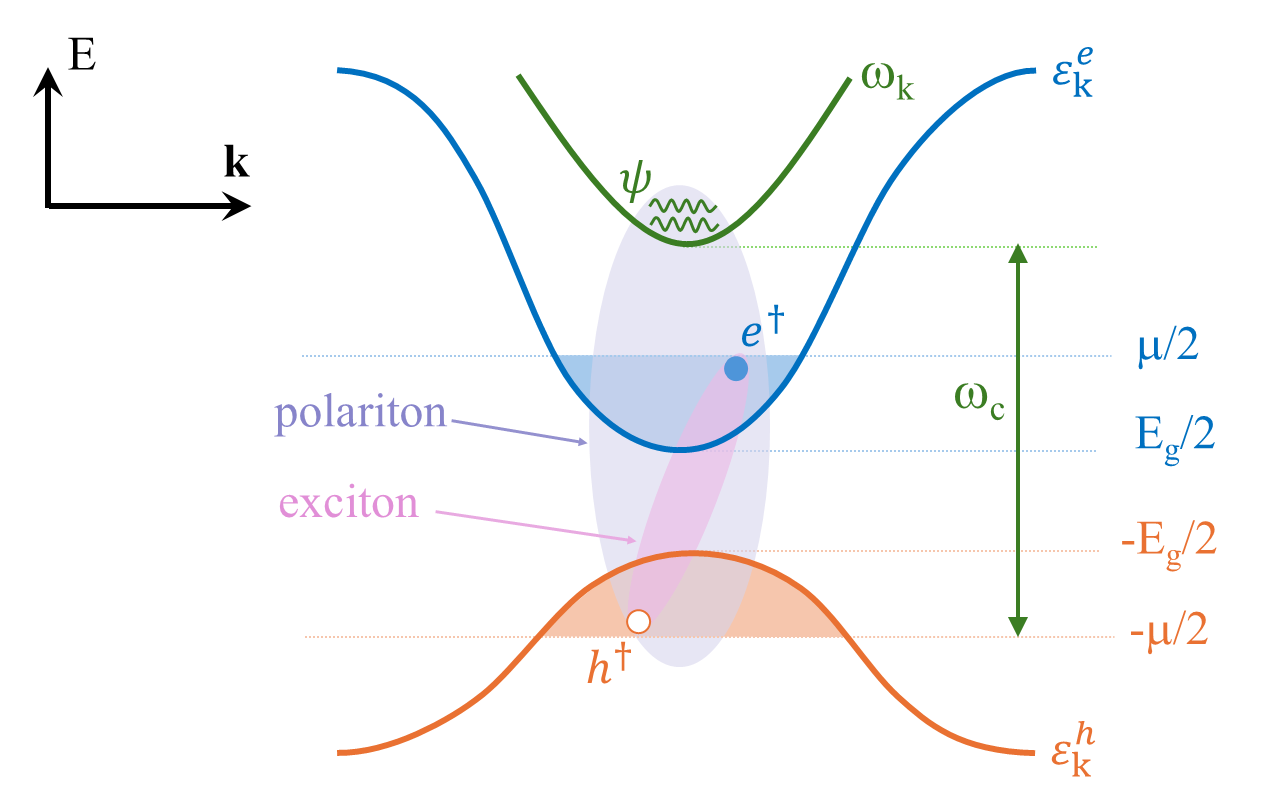}
\caption{Sketched band structure and relevant energy scales for the microscopic Hamiltonian in Eq.~\eqref{eq1} addressing the exciton-polariton formation in an optical microcavity.}
\label{fig0}
\end{figure}

In Eq.~\eqref{eq2}, $t^{e(h)}$ is the nearest-neighbor hopping amplitude of the electrons (holes) and $\gamma_{\mathbf{k}} = 2(\cos k_x + \cos k_y)$ for the tight-binding 2D systems. In the lattice, the hopping integral of a carrier is inversely approximate to its effective mass. The comparison of the hopping amplitude can release the difference in the effective mass of carriers in the system. In this work, without the loss of generality, we choose $t^e=1$ as the unit of energy, and thus $t^h$ might be used to specify the mass imbalance of the electron-hole system. In a realistic situation, the \cor{valence} hole is heavier than the conduction electron, thus $0\leq t^h\leq 1$. At $t^h=0$, the \cor{valence} hole is completely localized, the \cor{valence} band is flat and the effective mass of the holes is infinite. On the other hand, at $t^h=1$, the effective mass of the \cor{valence} holes is equal to that of the conduction electrons and the system is mass equivalent. $E_g$ in Eq.~\eqref{eq2} represents the energy difference between the bottom of the conduction electron band and the top of the valence hole band, it thus classifies semiconducting ($E_g > 0$) or semimetal ($E_g < 0$) state. $\omega_{c}$ in Eq.~\eqref{eq5} is a zero-point cavity frequency. In both Eqs.~\eqref{eq2} and \eqref{eq5}, $\mu$ is the chemical potential that would be used in the thermal dynamic equilibrium situation to adjust the total excitation density
\begin{equation}\label{eq3}
n = n_{}^{ph} + \frac{n_{}^e+n_{}^h}{2},
\end{equation}
where $n_{}^{ph} = \sum_{\mathbf{q}}\langle \psi_{\mathbf{q}}^\dagger \psi^{}_{\mathbf{q}} \rangle/N$ is density of photons, whereas $n_{}^{e} = \sum_{\mathbf{k}}\langle e^\dagger_{\bf{k}}e^{}_{\bf{k}}\rangle/N$ and $n_{}^{h} = \sum_{\mathbf{k}}\langle h^\dagger_{\bf{k}}h^{}_{\bf{k}}\rangle/N$ are the density of conduction electrons and that of \cor{valence} holes, respectively. $N$ is number of lattice sites. \cor{In this work, the photon chemical potential is set equal to the electronic ones to ensure the polariton condensate remains in equilibrium. Normally, photon chemical potential is zero because photons can be freely created or destroyed, but in confined systems like microcavities, it can be non-zero~\cite{PRL.105.056401}. Here, setting them equal likely helps balance the photonic (light) and excitonic (electron-hole pair) parts, making the system stable, especially as it shifts from being more like excitons to more like photons with higher excitation. Figure~\ref{fig0} illustrates the Hamiltonian for the consideration.}

The last two terms of the Hamiltonian in Eq.~\eqref{eq1} address the interactions in the electron-hole-photon in microcavity with $U$ as the localized electron-hole Coulomb attraction strength and $g$ as the light-matter coupling constant. Without the Coulomb interaction, the Hamiltonian given in Eq.~\eqref{eq1} delivers the Dicke model originally proposed to describe the coherent coupling of an ensemble with two-level atoms to single-mode light field~\cite{PR.93.99}, and it has been widely used to examine the polariton condensate in optical microcavity~\cite{PRB.64.235101,PRL.96.230602,RMP.82.1489,RMP.85.299}. In the meanwhile, if only the Coulomb attraction between the conduction electrons and \cor{valence} holes is considered, the Hamiltonian in Eq.~\eqref{eq1} turns into the extended Falicov-Kimball model~\cite{BGBL04,Ba02b}. In its framework, formation and stability of the excitonic condensates in semimetal/semiconducting materials have been profoundly investigated~\cite{IPBBF08,PRB.107.115106,PRB.109.085105,PRB.110.235143}. In this work, both the Coulomb interaction in the extended Falicov-Kimball model and the matter-light coupling in the Dicke idea are considered on an equal footing in the compact Hamiltonian. In this aspect, probably natural signatures of the polariton condensates and especially its competition to the excitonic and the photonic condensation states in a microcavity, therefore, would be delivered.

\subsection{Unrestricted Hartree-Fock approximation}

The model proposed in Eq.~\eqref{eq1} is the many-particle Hamiltonian and in general it is not exactly solvable and one needs to find some approximations. In this work, the Hamiltonian is solved by using the unrestricted Hartree-Fock approximation. The UHFA is an extended version of the Hartree-Fock approximation in which all possibly off-diagonal couplings are considered. In the ground state, the UHFA has proved as an applicable approach treating to a strongly correlated electronic systems as done from other effective methods, for instance, the dynamical mean-field theory~\cite{Georges06,Cz99} or the density matrix renormalization group~\cite{PRB.95.045101,PRB.102.205111,PRX.10.031034}. \cor{In this work, we are concerned on the instability of the polaritonic condensation states with the formation of the excitonic and photonic coherence. In the situation, the off-diagonal hybridization between the conduction electrons and valence holes and photonic polarization are considered. By eliminating all fluctuation parts, the Hamiltonian given in Eq.~\eqref{eq1} can be written as}
\begin{align}\label{eq6}
\mathcal{H}_\textrm{UHF} =&\sum_{\bf{k}}{\bar{\varepsilon}^e_{\bf{k}} e^\dagger_{\bf{k}} e_{\bf{k}}} + \sum_{\bf{k}}{\bar{\varepsilon}^h_{\bf{k}} h^\dagger_{\bf{k}} h^{}_{\bf{k}}}+ \Delta \sum_{\bf{k}}{( e^\dagger_{\bf{k}}h^\dagger_{\bf{-k}}+\textrm{H.c.})}\nonumber\\
&+\sum_{\bf{q}}\omega^{}_{\bf{q}} {\psi^\dagger_{\bf{q}} \psi^{}_{\bf{q}}} +( \sqrt{N}\Gamma\psi^\dagger_0+\textrm{H.c.}).
\end{align}
Here, we have simplified the interaction operators in Eq.~\eqref{eq1} by depleting all fluctuation parts and constants, such that 
\begin{align}
 e_{\mathbf{k}_1+\mathbf{q}}^\dagger e^{}_{\mathbf{k}_1} h_{\mathbf{k}_2-\mathbf{q}}^\dagger h^{}_{\mathbf{k}_2}\approx &\delta_{{\bf q},0}(n^e_{{\bf k}_1} h_{\mathbf{k}_2}^\dagger h^{}_{\mathbf{k}_2} + n^h_{{\bf k}_2} e_{\mathbf{k}_1}^\dagger e^{}_{\mathbf{k}_1})\nonumber\\
 &+ \delta_{{\bf k}_1,-{\bf k}_2}(d^{}_{{\bf k}_1}e^{\dagger}_{\mathbf{k}_1+{\bf q}}h^{\dagger}_{-\mathbf{k}_1-{\bf q}}+\textrm{H.c.}),
\end{align}
and
\begin{align}
e_{\mathbf{k}+\mathbf{q}}^\dagger h_{-\mathbf{k}}^\dagger \psi^{}_{\mathbf{q}} \approx \delta_{{\bf q},0} (d_{\mathbf{k}}\psi^{}_{\mathbf{q}}+\langle  \psi^{}_{\mathbf{q}}\rangle e_{\mathbf{k}}^\dagger h_{-\mathbf{k}}^\dagger),
\end{align}
where, the denotations $n^e_{{\bf k}}=\langle e_{\mathbf{k}}^\dagger e^{}_{\mathbf{k}}\rangle$, $n^h_{{\bf k}}=\langle h_{\mathbf{k}}^\dagger h^{}_{\mathbf{k}}\rangle$, and $d_{\bf{k}}=\langle e^\dagger_{\bf{k}}h^\dagger_{\bf{-k}}\rangle=\langle h^{}_{\bf{-k}}e^{}_{\bf{k}}\rangle$ have been used. 

In the effective Hamiltonian written in Eq.~\eqref{eq6}, the electronic excitation energies $\bar{\varepsilon}^{e(h)}_{\bf{k}}$ have involved Hartree shifts
\begin{eqnarray}\label{eq7}
\bar{\varepsilon}^{e(h)}_{\bf{k}} = \varepsilon^{e(h)}_{\bf{k}} -Un_{}^{h(e)},
\end{eqnarray}
and additional fields act as the order parameters of the exciton-polariton condensation
\begin{equation}\label{eq8}
\Delta=-\frac{g}{\sqrt{N}}\langle\psi_0\rangle-\frac{U}{N}\sum_{\bf{k}}d_{\bf{k}},
\end{equation}
\begin{equation}\label{eq9}
\Gamma=-\frac{g}{N}\sum_{\bf{k}}d_{\bf{k}}.
\end{equation}
In Eq.~\eqref{eq8}, $d_{\bf k}$ plays the role of the wave-vector-resolved intensity of the electron-hole pair amplitude that indicates the coherent formation of electron-hole pairs in the excitonic condensate. Its non-zero value specifies a macroscopic occupation of excitonic states and embodies the essential characteristics of phase coherence and collective behavior in the system. $\Delta_\textrm{ex}=U\sum_{\bf{k}}d_{\bf{k}}/N$ thus can be called the excitonic condensate order parameter. In the meanwhile, $\langle\psi_0\rangle$ gives insight into the average amplitude of the electromagnetic field, which relates to the coherent state of photons. In this sense, we denote $\Delta_{\textrm{ph}}=g\langle\psi_0\rangle/\sqrt{N}$ as the photonic condensate order parameter. Apparently, $\Delta_{\textrm{ph}}$ plays a crucial point releasing the quantum coherent states in the system due to the matter-light coupling. The energy gap given in Eq.~\eqref{eq8} depends both on the excitonic and photonic coherences in the system. That indicates the energy bound of the polariton condensate, a superposition of the electron-hole pair and photon coherent state. Without the matter-light coupling, i.e., $g=0$, one finds only the excitonic condensation if $\Delta$ is nonzero. Depending on which $\Delta_\textrm{ex}$ or $\Delta_{\textrm{ph}}$ is dominated, the system can stabilize in the polariton condensate with either excitonic-like or photonic-like \cor{characterization}~\cite{Kamide2011}. The polariton condensate is specified if the two order parameters are comparable. \cor{Note here that, in general, the polariton condensate is found typically at lower polariton (LP) mode in the dispersion of the strong coupling between cavity photons and excitons. In order to characterize the polariton composition, one uses the Hopfield coefficients $|C|^2$ and $|X|^2$ that give the photonic and excitonic fractions of a single LP mode, respectively~\cite{Im20}. In our definition, $\Delta_\textrm{ph}$ scales with $\langle\psi_0\rangle$ which is proportional to the photon number $n_\textrm{ph} = |C|^2n_\textrm{pol}$ and $\Delta_\textrm{ex}$ scales with $\sum_{\bf k} d_{\bf k}$ where the excitonic pairing amplitude approximates to $n_\textrm{ex}= |X|^2n_\textrm{pol}$, $n_\textrm{pol}$ here is the LP polariton number in the condensate. In this sense, one finds the excitonic fraction $\Delta_\textrm{ex}/\Delta\sim |X|^2$ and photonic fraction $\Delta_\textrm{ph}/\Delta\sim |C|^2$ that align to the common definitions in the standard polariton theory~\cite{Im20}, but here they are applied to order parameters rather than wavefunction weights.} The expression in Eq.~\eqref{eq8} reveals that the electron-hole coherence in the microcavity depends on both the Coulomb attraction and the matter-light coupling. The mutual influence of these interactions on the stability of the condensation states in the system thus will be of interest. 

The Hamiltonian in Eq.~\eqref{eq6} can be divided into the electronic (the first line) and photonic (the rest ones) parts which can be separately diagonalized. Indeed, using the Bogoliubov transformation by defining new fermionic operators
\begin{eqnarray}\label{13}
c^\dagger_{1\bf{k}}&=& u_{\bf{k}} e^\dagger_{\bf{k}}+v_{\bf{k}} h^{}_{-\bf{k}},\nonumber\\
c^\dagger_{2\bf{k}}&=& -v_{\bf{k}} e^\dagger_{\bf{k}}+u_{\bf{k}} h^{}_{-\bf{k}},
\end{eqnarray}
with the prefactors
\begin{eqnarray}\label{13a}
u^2_{\bf{k}} = \frac{1}{2}\left[ 1 + \textrm{sgn}( \bar{\varepsilon}^e_{\bf{k}} + \bar{\varepsilon}^h_{\bf{k}}) \frac{ \bar{\varepsilon}^e_{\bf{k}} + \bar{\varepsilon}^h_{\bf{k}}}{W_{\bf{k}}}\right] ,\\
v^2_{\bf{k}} = \frac{1}{2}\left[ 1 - \textrm{sgn}( \bar{\varepsilon}^e_{\bf{k}} + \bar{\varepsilon}^h_{\bf{k}}) \frac{ \bar{\varepsilon}^e_{\bf{k}} + \bar{\varepsilon}^h_{\bf{k}}}{W_{\bf{k}}}\right] \label{13b},
\end{eqnarray}
and
\begin{equation}\label{14}
W_{\bf{k}} = \sqrt{(\bar{\varepsilon}^e_{\bf{k}} + \bar{\varepsilon}^h_{\bf{k}} )^2 + 4|\Delta| ^2}.
\end{equation}
then one arrives at a diagonalized form of the electronic part
\begin{equation}
\mathcal{H}^\textrm{e}_\textrm{UHF}=\sum_{\alpha{\bf k}}E^\alpha_{\bf k}c^\dagger_{\alpha\bf{k}}c^{}_{\alpha\bf{k}},  
\end{equation}
where
\begin{eqnarray}\label{9}
E^\alpha_{\bf{k}} = \frac{\bar{\varepsilon}^e_{\bf{k}} - \bar{\varepsilon}^h_{\bf{k}}}{2} -(-1)^\alpha \frac{\textrm{sgn}(\bar{\varepsilon}^e_{\bf{k}} + \bar{\varepsilon}^h_{\bf{k}})}{2}W_{\bf{k}}.
\end{eqnarray}
The photonic part can also be diagonalized by defining new bosonic operator
\begin{equation}\label{10}
\Psi^\dagger_{\bf{q}} = \psi^\dagger_{\bf{q}} + \sqrt{N}\left(\Gamma/\omega_0\right)\delta_{\bf{q,0}}\,.
\end{equation}
It then becomes
\begin{equation}\label{11}
\mathcal{H}^\textrm{ph}_\textrm{dia}=\sum_{\bf{q}}\omega_{\bf q}{\Psi^\dagger_{\bf{q}} \Psi_{\bf{q}}}.  
\end{equation}
Once the whole parts in the effective Hamiltonian in Eq.~\eqref{eq6} are diagonalized, we can simply evaluate the expectation values to close a set of self-consistent equations
\begin{align}\label{15}
&n^e_{\bf{k}} = u^2_{\bf{k}} f(E^1_{\bf{k}})  + v^2_{\bf{k}}f(E^2_{\bf{k}}),\\ 
&n^h_{\bf{k}} = 1-v^2_{\bf{k}} f(E^1_{\bf{k}})  -u^2_{\bf{k}}f(E^2_{\bf{k}}),\\ 
&d_{\bf{k}}  = [f(E^1_{\bf{k}}) - f( E^2_{\bf{k}})] \textrm{sgn} (E^1_{\bf{k}} - E^2_{\bf{k}}) \frac{\Delta}{W_{\bf{k}}},\label{16}\\
&\langle \psi^{}_{0}\rangle =-\frac{\sqrt{N}\Gamma}{\omega_0}.\label{18}
\end{align}
Here we have defined $f(E)=1/(e^{\beta E}+1)$ as the Fermi-Dirac distribution function with $\beta = 1/T$ being  {the inverse of the temperature $T$}. Note that the total excitation density in Eq.~\eqref{eq3} is determined only if the density of photon is specified. From the diagonalized form of the photonic part in Eq.~\eqref{11}, one finds
\begin{equation}\label{19}
n^\textrm{ph}_{\bf q}=\langle \psi_{\mathbf{q}}^\dagger \psi^{}_{\mathbf{q}} \rangle=p(\omega_{\bf q})+ \frac{N\Gamma^2}{\omega^2_0}\delta_{\bf{q,0}}, 
\end{equation}
where $p(\omega)=1/(e^{\beta \omega}-1)$ is the Bose-Einstein distribution function.

From Eqs.~(\ref{eq7})-(\ref{eq9}) and (\ref{15})-(\ref{18}), we find a set of self-consistent equations that might be solved numerically. In the features of the UHFA, the physical signatures of the electron-hole-photon system modelled in the Hamiltonian~\eqref{eq1} would be delivered.

\subsection{Dynamical susceptibilities}

In order to analyze the stability of the polariton condensation states, inspecting profoundly the fluctuations of the excitonic and photonic coherences is also an essential task. In this work, the fluctuations are examined in the signatures of the excitonic and photonic dynamical susceptibility functions. Defining $X^\dagger_{\bf q}=\sum_{\bf k}e^\dagger_{{\bf k}+{\bf q}}h^\dagger_{-\bf k}/\sqrt{N}$ as a creation operator of the exciton with momentum ${\bf q}$, the linear response theory leads us to an expression of the excitonic dynamical susceptibility function like
\begin{eqnarray}\label{eq13}
\chi^\textrm{ex}({\bf{q}},\omega)=-\frac{1}{N}\sum_{\bf{k},\bf{k'}} \langle\langle h^{}_{-\mathbf{k}}e^{}_{\bf{k}+\bf{q}}; e^\dagger_{\bf{k'}+\bf{q}}h^\dagger_{-\bf{k'}}\rangle\rangle_\omega.
\end{eqnarray}
In the same way, we can define the photonic dynamical susceptibility function as
\begin{equation}\label{eq14}
\chi^\textrm{ph}({\bf{q}},\omega)=-\langle\langle \psi^{}_{\bf{q}}; \psi^\dagger_{\bf{q}}\rangle\rangle_\omega.
\end{equation}
In the above equations, the symbol $\langle\langle \cal{A};\cal{A}^\dagger \rangle\rangle_\omega$ denotes the retarded Green’s function of the bosonic particle with the annihilation operator ${\cal A}$. The correlation function can be evaluated based on the equation of motion (EOM) method, i.e., 
\begin{equation}\label{14a}
\omega\langle\langle \cal{A};\cal{A}^\dagger \rangle\rangle_\omega=\langle[\cal{A},\cal{A}]\rangle+ \langle\langle [\cal{A},\mathcal{H}];\cal{A}^\dagger \rangle\rangle_\omega.
\end{equation}
Using the EOM method above the last term in Eq.~\eqref{14a} would lead to higher order Green's functions. In order to derive an analytical expression of the correlation function, one needs some truncations. In our target there, we use the random phase approximation (RPA) and one straightforwardly arrives, for instance, a solution of the excitonic susceptibility function
\begin{equation}\label{eq17}
\chi^\textrm{ex}({\bf{q}},\omega)=-\frac{1}{[\chi^\textrm{ex}_0({\bf{q}},\omega)]^{-1}+U-{g^2}\chi^\textrm{ph}_0({\bf{q}},\omega)},
\end{equation}
where $\chi^\textrm{ex}_0({\bf{q}},\omega)$ and $\chi^\textrm{ph}_0({\bf{q}},\omega)$ are the bare excitonic and photonic susceptibility functions, that, respectively, read
\begin{equation}\label{eq19}
\chi^\textrm{ex}_0({\bf{q}},\omega)=\frac{1}{N}\sum_{\bf{k}}\frac{1- n^h_{\bf{k}}-n^e_{{\bf{k}}+\bf{q}}}{\omega-\bar{\varepsilon}^{e}_{{\bf{k}}+{\bf{q}}}-\bar{\varepsilon}^{h}_{\bf{k}}},
\end{equation}
and
\begin{equation}\label{eq20}
\chi^\textrm{ph}_0({\bf{q}},\omega) =\frac{1}{\omega-\omega_{\bf{q}}}. 
\end{equation}

In the same manner, we also obtain an analytical expression of the photonic dynamical susceptibility function. It is written as
\begin{equation}\label{eq18}
\chi^\textrm{ph}({\bf{q}},\omega)=\frac{1}{[\chi^\textrm{ph}_0({\bf{q}},\omega)]^{-1}-g^2\chi^\textrm{ex}_0({\bf{q}},\omega)/[1+U\chi^\textrm{ex}_0({\bf{q}},\omega)]}.
\end{equation}
where $\chi^\textrm{ex}_0({\bf{q}},\omega)$ and $\chi^\textrm{ph}_0({\bf{q}},\omega)$ are given in Eqs.~(\ref{eq19}) and (\ref{eq20}). 

The expression of excitonic and photonic susceptibility functions involves the presence of both the Coulomb interaction and matter-light coupling. At $g=0$, apparently, one delivers the regular expression of the excitonic susceptibility function once only the interlayer Coulomb interaction is taken into account, whereas, the photonic susceptibility function returns to that solution of the free photon system.

\section{Numerical results and discussion}
In order to examine profoundly the stability of the polariton condensation states in a microcavity, the set of self-consistent equations above in  Eqs.~(\ref{eq7})-(\ref{eq9}) and (\ref{15})-(\ref{18}) would be solved numerically. Its solutions release signatures of the quasi-particle energies, polariton condensate order parameter and also the excitonic and photonic susceptibilities. These issues would address the fluctuations and stabilities of the condensation states in the system. To process the numerical calculation we consider the electrons, holes, and photons in a 2D system with $N=200\times 200$ lattice sites. In a microcavity with a given cavity frequency $\omega_c$, the system can be semimetal or semiconducting state depending on the tuning parameter $d$ following the relation that $d=\omega_{c}-E_g$~\cite{Kamide2011}. If this parameter is small, the excitation energy of the electron-hole pair is near the bottom of the photon energy band, making the photonic contribution be significant. Conversely, if the detuning parameter is large, the photonic band overlaps less with the electron and hole bands, and the photon contribution to the exciton-polariton condensate becomes less significant~\cite{Kamide2011,NHAM2017}. In this work, we fix $\omega_{c} = 0.5$ and the matter-light coupling $g=0.2$ in order to concentrate on examining the condensate stabilities in the influence of the electron-hole mass imbalance, Coulomb interaction and detuning in the ground state. Indeed, varying the matter-light coupling and/or the bottom of the photon band only monotonously changes the amplitude of the pairing states in both small and large detunings~\cite{PBF16}. Here, the chemical potential $\mu$ is adjusted to specify the excitation density $n$ given in  Eq.~\eqref{eq3}. 

\subsection{Condensate order parameters}
\begin{figure}[hbt]
\includegraphics[width=0.470\textwidth]{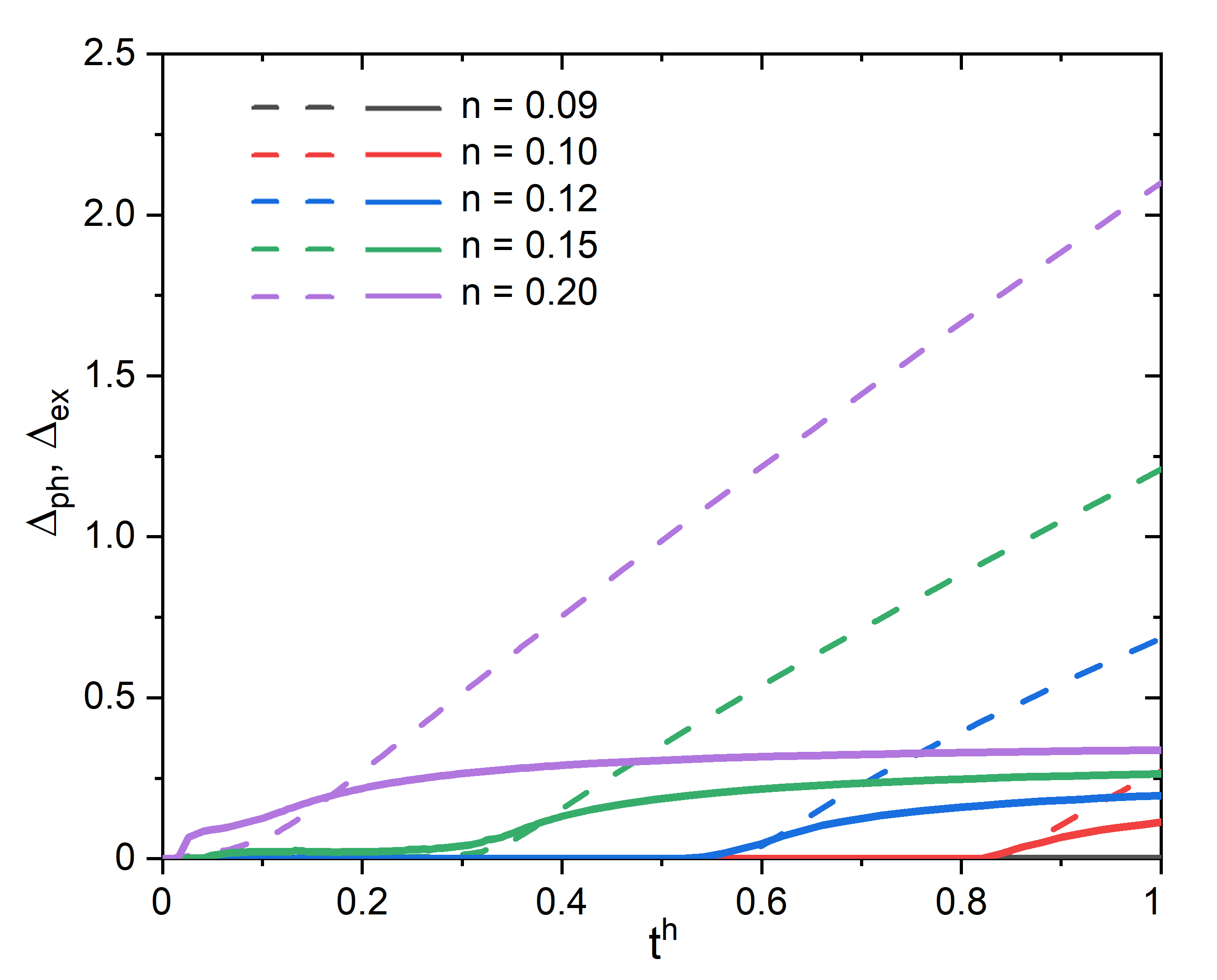}
\caption{The excitonic $\Delta_{ex}$ (solid lines) and the photonic $\Delta_{ph}$ (dashed lines) condensate order parameters as a function of the hole hopping integral $t^h$ for different values excitation density $n$ at Coulomb interaction $U=1$ and detuning $d=2$.}
\label{fig1}
\end{figure}

To discuss the impact of the mass imbalance on the stability of the condensates in the system, in the first, we address in Fig.~\ref{fig1} a behavior of the excitonic ($\Delta_\textrm{ex}$) and photonic ($\Delta_\textrm{ph}$) condensate order parameters as functions of the hole hopping amplitude $t^h$ at Coulomb interaction $U=1$ and detuning $d=2$ for some excitation densities $n$. For a given excitation density, as increasing $t^h$ or lowering the mass imbalance we always find an enhancement of the excitonic and photonic condensate order parameters. Indeed, once $t^h$ is small or the mass of holes is large in comparison to that of electrons. The holes seem to be inflexible and they tend to be localized, that restricts the possibility of the hybridization between the electrons in the conduction band and holes in the \cor{valence} band. The excitons thus are hardly formed. Apparently at $t^h=0$, the hole \cor{valence} band is completely flat or the mass of holes is infinite, one finds that both $\Delta_\textrm{ex}$ and $\Delta_\textrm{ph}$ are zero. In that feature, at low excitation densities, Fig.~\ref{fig1} shows us that the condensates of excitons and photons in microcavity occur only if the $t^h$ is sufficiently large. \cor{The enhancement of coherent stability with a reduction in mass imbalance illustrated in Fig.\ref{fig1} aligns qualitatively with the experimental observations reported in Ref.~\cite{PRB.93.115205}. This suggests that ZnO exhibits more stable polariton condensates when compared to GaN. Note here that the mass imbalance of electrons and holes in ZnO [$\sim 0.53$~\cite{JPC.100.1027}] is larger than that in GaN [$\sim 0.16$~\cite{Suzuki1995}].} Increasing the excitation density, the critical value of $t^h$ for the formation of the condensates shifts to the left corresponding to a lower value of the hole hopping integral. In the case of large excitation density, one also finds a saturation of $\Delta_\textrm{ex}$ whereas $\Delta_\textrm{ph}$ continuously increases as enlarging $t^h$. Indeed, as increasing $t^h$, the number of electron-hole pairs increases until the hole \cor{valence} band reaches the photon level and thereafter the photonic excitations replace those of the electron-hole pairs to minimize the total energy~\cite{Kamide2011}. 

\begin{figure}[hbt]
\includegraphics[width=0.490\textwidth]{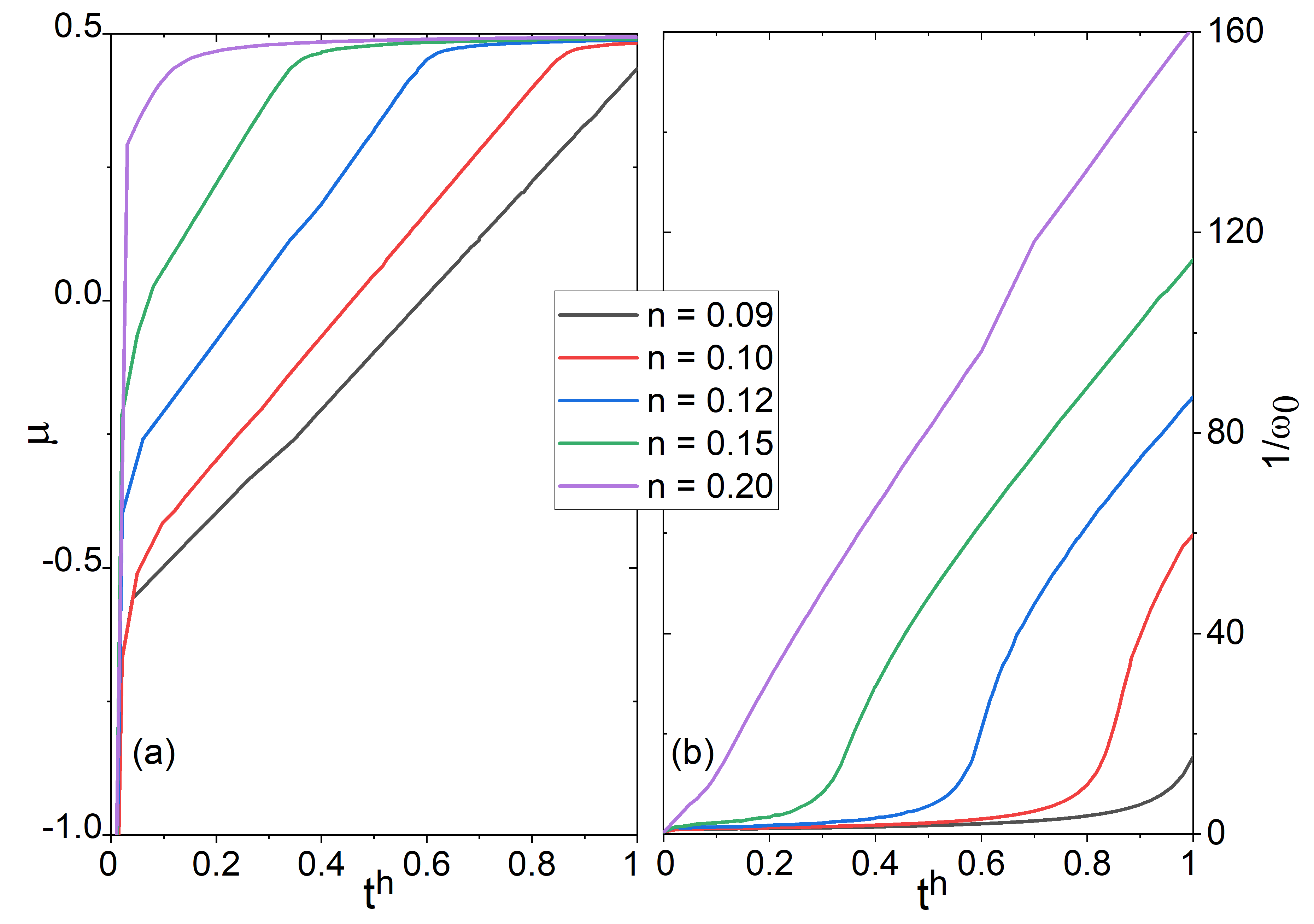}
\caption{The chemical potential $\mu$ and the inverse of the photon excitation energy band bottom $1/\omega_{0}$ versus the hole hopping integral $t^h$ for the sets of parameters mentioned in Fig.~\ref{fig1}.}
\label{fig2}
\end{figure}

At any given set of model parameters, $\Delta_\textrm{ex}$ and $\Delta_\textrm{ph}$ are mutually related such that $\Delta_\textrm{ph}=\Delta_\textrm{ex}g^2/(U\omega_0)$. The relation can be derived from the definitions of the condensate order parameters below Eq.~\eqref{eq8} with the help of Eqs.~\eqref{eq9} and~\eqref{18}. Apparently, for fixed interaction strengths $g$ and $U$, a ratio $\Delta_\textrm{ph}/\Delta_\textrm{ex}$ or the competition of the excitonic- and photonic-like polariton condensates depends only on $\omega_0$, the bottom of the photonic excitation energy band. Once the excitation density is small, the chemical potential locates slightly above the bottom of the conduction electron band or slightly below the top of the \cor{valence} hole band. In that situation, as long as the \cor{valence} hole band is narrow, the chemical potential would settle far from the zero-point cavity frequency $\omega_c$ and the photonic contribution then is neglectable. The photons tend to play their role contributing to the total excitation density and the coherent states only if the mass imbalance becomes sufficiently small [see Fig.~\ref{fig2}(a)]. Indeed, by this way, the top of the hole \cor{valence} band might reach the photonic bottom band and the hybridization between excitons and photons or the polariton condensate happens. Note here that both $\Delta_\textrm{ex}$ and $\Delta_\textrm{ph}$ mutually coexist and thus the polariton condensate occurs simultaneously with the excitonic and photonic condensates. However, depending on the ratio $\Delta_\textrm{ph}/\Delta$ or $\Delta_\textrm{ex}/\Delta$, one can characterize the polariton condensate as either the excitonic-like or photonic-like condensate if $\Delta_\textrm{ph}/\Delta<20\%$ or $\Delta_\textrm{ph}/\Delta>80\%$, respectively~\cite{PRL.105.056401,Kamide2011}. In the case of $n=0.1$ (red lines in Fig.~\ref{fig1}), $\Delta_\textrm{ph}$ becomes dominated to $\Delta_\textrm{ex}$ suddenly as they are nonzero because the chemical potential rapidly reaches the zero-point cavity frequency or $1/\omega_0$ becomes large [cf. Fig.~\ref{fig2}(b)]. Increasing the excitation density enriches the possibility of the hybridization between electrons and holes. The excitons thus can be formed and condensed in situations with larger mass imbalance. However, due to being stabilized even at small $t^h$, the polaritons are formed with less signature of the photons because they are located a bit above the chemical potential and one finds the excitonic-like polariton condensate. The photons only become significantly important in the case of sufficiently large $t^h$ once the top of the \cor{valence} band approaches the bottom of the photon band. The mobility of the holes in the situations becomes comparable to that of electrons. By lowering the mass imbalance one finds polariton and then photonic-like polariton condensation states. The complicated competition of $\Delta_\textrm{ex}$ and $\Delta_\textrm{ph}$ in varying the mass imbalance opens an enrichment phase structure in the microcavity. The detailed phase diagram of the coherent states in the system would be left to be discussed later.

\begin{figure}[hbt]
\includegraphics[width=0.50\textwidth]{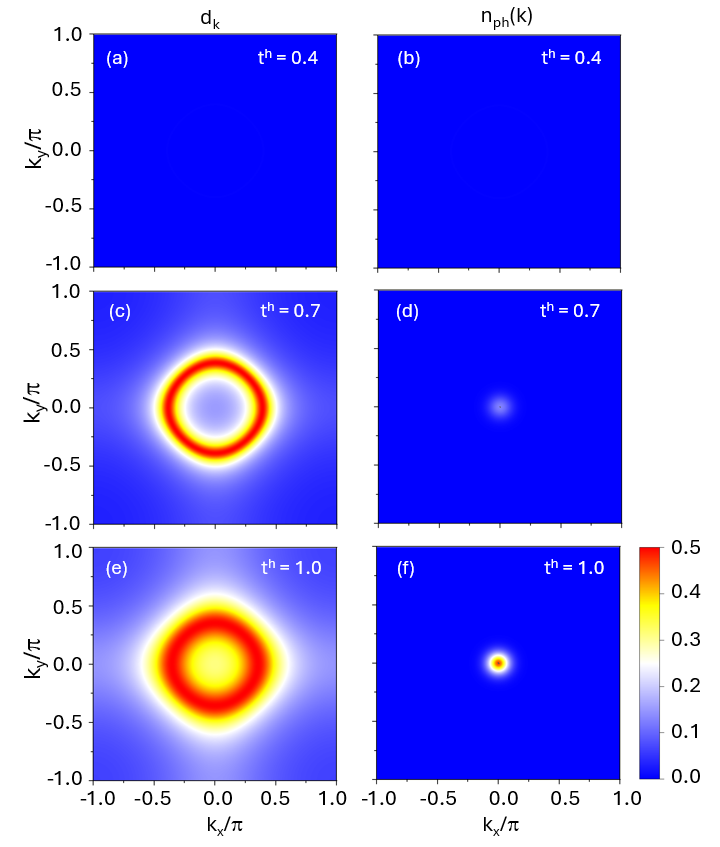}
\caption{Electron-hole pair amplitude $d_{\bf k}$ (left panels) and the photonic density $n_\textrm{ph}({\bf k})$ (right panels) in the first Brillouin zone for three different values of $t^h$ at $U=1$, $d=2$, and $n=0.12$.}
\label{fig3}
\end{figure}

In order to analyze the nature of the phase structures in the mass anisotropic electron-hole system with the photon in microcavity we discuss below the momentum distribution of the electron-hole pair amplitude $d_{\bf k}$ and the photonic density $n_\textrm{ph}({\bf k})$, they are respectively defined in Eqs.~\eqref{16} and~\eqref{19}. In Fig.~\ref{fig3} we address wave-vector-resolved intensity of $d_{\bf k}$ and $n_\textrm{ph}({\bf k})$ for the typical set of parameters in Fig.~\ref{fig1} at $n=0.12$ in the first Brillouin zone. For the given value of the excitation density, Fig.~\ref{fig3} shows us that once the hole integral hopping $t^h=0.4$, both $d_{\bf k}$ and $n_\textrm{ph}({\bf k})$ are zero in the whole first Brillouin zone [see Fig.~\ref{fig3}(a) and~\ref{fig3}(b)], indicating that the system is out of the condensation states. Both the excitonic and photonic condensate order parameters are zero (cf. blue lines in Fig.~\ref{fig1}). Nonzero values of $d_{\bf k}$ and $n_\textrm{ph}({\bf k})$ are found only if $t^h$ is sufficiently large (see the left panels). Nonzero values of $d_{\bf k}$ at a certain momentum indicate the possibility of the hybridization between electrons in the conduction band and holes in the \cor{valence} band, specifying the stability of the excitonic condensate. In the whole range of the $t^h$, we always find the two-pick structure of the $d_{\bf k}$ or the excitonic condensate stabilizes in the BCS-type. The peaks locate at the Fermi momentum at which the two bands intersect each other. Note here that the system is settled by fixing the excitation density. In the case of small $t^h$ or large mass imbalance, the excitonic level is far from the bottom of the photon band and the excitation density is mainly contributed due to the electron-hole plasma. The two bands thus intersect each other even once the \cor{valence} band is nearly flat. Increasing $t^h$ also broadens the spectrum in the momentum space and the peaks in $d_{\bf k}$ become smeared out. Indeed, by increasing $t^h$ or lowering the mass imbalance, the holes tend to be flexible which reinforces the coherence between the electrons and holes even they deviate from the Fermi level. More electrons and holes with momenta around the Fermi ones might couple each other to form the excitons, enlarging the excitonic condensate order parameter. In the situation with more flexibility of the holes, the broadening of the electron-hole pair spectrum in momentum space also indicates the reduction of the size of the excitons in the real space. In the case of sufficiently large $t^h$, the electron-hole hybridization might reach the bottom of the photon excitation band and the matter-light coupling also plays roles in establishing the coherence of electron-hole pairs. The excitonic-like polariton and then the polariton condensates are reinforced. That signature can also be addressed in the photon distribution $n_\textrm{ph}({\bf k})$ in the right panels of Fig.~\ref{fig3}. From the Eq.~\eqref{19} one finds $n_\textrm{ph}({\bf k})$ capturing a sharp peak at zero momentum specifying the photonic coherent state if the excitonic condensate is risen up. Indeed, as increasing the hole mobility, fixing the excitation density brings the electron-hole excitation to that of the photon band. Due to the matter-light coupling, the polaritonic coherence is established. The peaks in the photon distribution $n_\textrm{ph}({\bf k})$ become rapidly sharp once the Fermi level or the hybridization excitation reaches closely the bottom of the photon band. That happens if the masses of electrons and holes respectively tend to be identical [cf.~Fig.~\ref{fig3}(f)]. 

\subsection{Spectral functions}

In order to gain further insight into the mass-imbalance influence on the coherent states in microcavity, we address here the wave vector- and frequency-resolved single-particle spectral functions in varying the hopping integral $t^h$. The single particle spectral functions measure the photoemission spectrum of particles with wave vector ${\bf k}$, then the occupied densities of single-particle in the quantum coherence states would be addressed. The single particle spectral function for the electrons can be defined as
\begin{equation}
A^e({\bf k},\omega)=-\frac{1}{\pi}\textrm{Im}G^e({\bf k},\omega),  
\end{equation}
where $G^e({\bf k},\omega)=\langle\langle e^{}_{\bf k};e^{\dagger}_{\bf k}\rangle\rangle_\omega$ is the retarded Green function of the conduction electrons. Similar to evaluating the dynamical susceptibility functions in the previous section, one delivers
\begin{equation}
A^e({\bf k},\omega)=u^2_{\bf k}\delta(\omega-E^1_{\bf k})+v^2_{\bf k}\delta(\omega-E^2_{\bf k}).
\end{equation}
Here, the prefactors $u^2_{\bf k}$ and $v^2_{\bf k}$ are defined in Eqs.~(\ref{13a}) and (\ref{13b}) and the quasi-particle excitation energies are given in Eq.~\eqref{9}. In the same way, we can find the spectral function of holes $A^h({\bf k},\omega)$ and photons $A^p({\bf k},\omega)$, respectively, as follows
\begin{eqnarray}
A^h({\bf k},\omega)&=&v^2_{\bf k}\delta(\omega+E^1_{\bf k})+u^2_{\bf k}\delta(\omega+E^2_{\bf k}),  \\
A^p({\bf k},\omega)&=&\delta(\omega-\omega_{\bf k}).
\end{eqnarray}
In the last equation, the quasi-particle energy of photons $\omega_{\bf k}$ is given in Eq.~\eqref{eq5}.

\begin{figure}[t]
\includegraphics[width=0.23\textwidth]{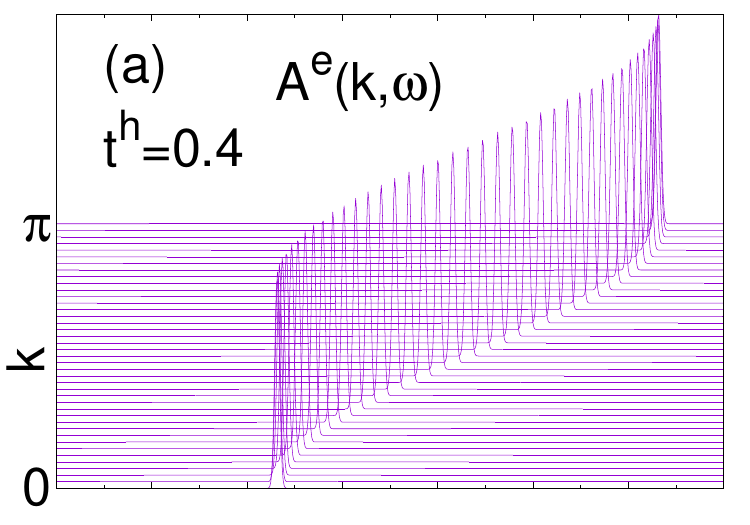}
\includegraphics[width=0.23\textwidth]{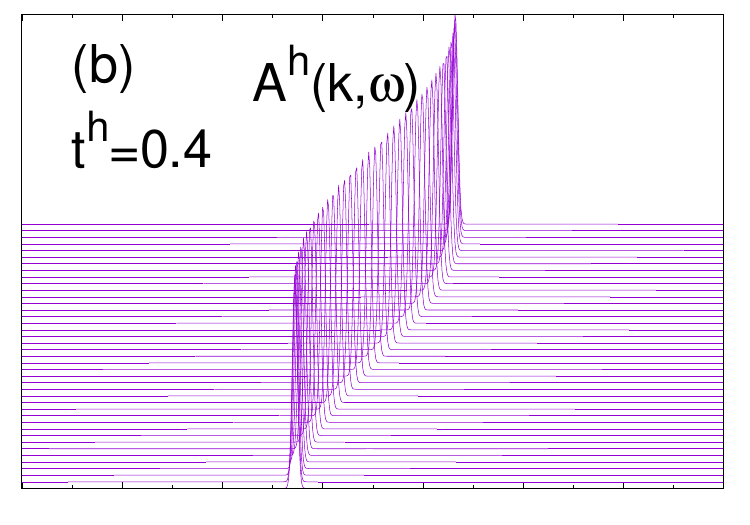}\\
\includegraphics[width=0.23\textwidth]{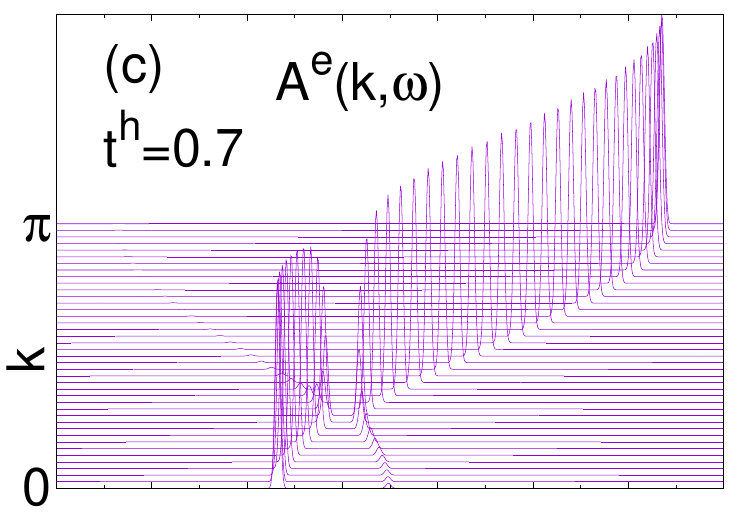}
\includegraphics[width=0.23\textwidth]{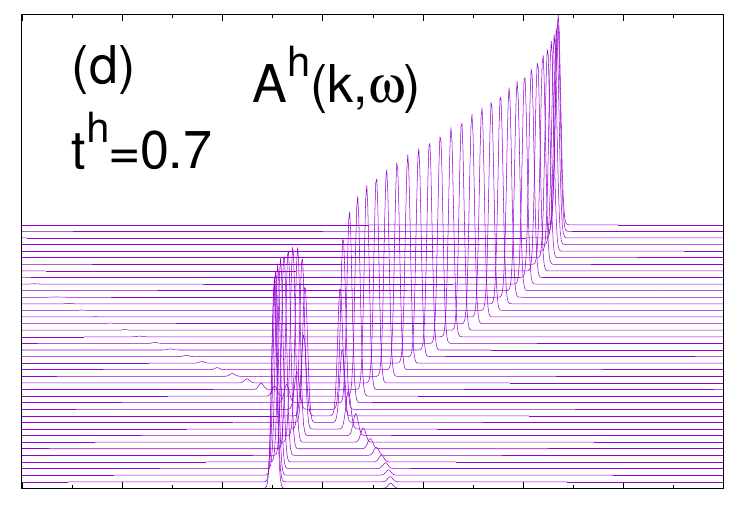}\\
\includegraphics[width=0.23\textwidth]{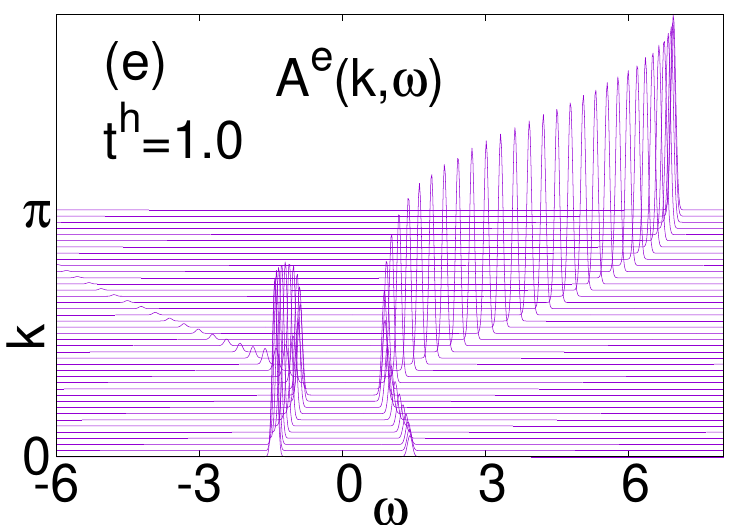}
\includegraphics[width=0.23\textwidth]{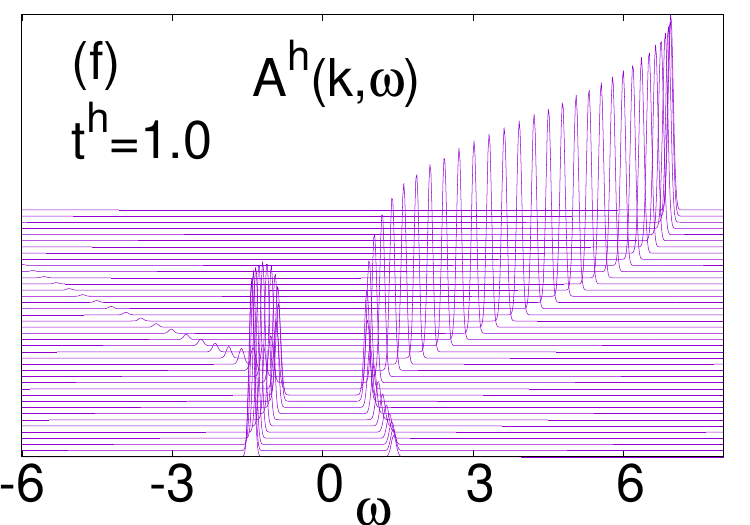}
\caption{Wave-number resolved photoemission spectra of electrons $A^e({\bf k},\omega)$ (left panels) and holes $A^h({\bf k},\omega)$ (right panels) for the parameters set in Fig.~\ref{fig3}.}
\label{fig4}
\end{figure}

\begin{figure*}[htb]
\includegraphics[width=0.275\textwidth]{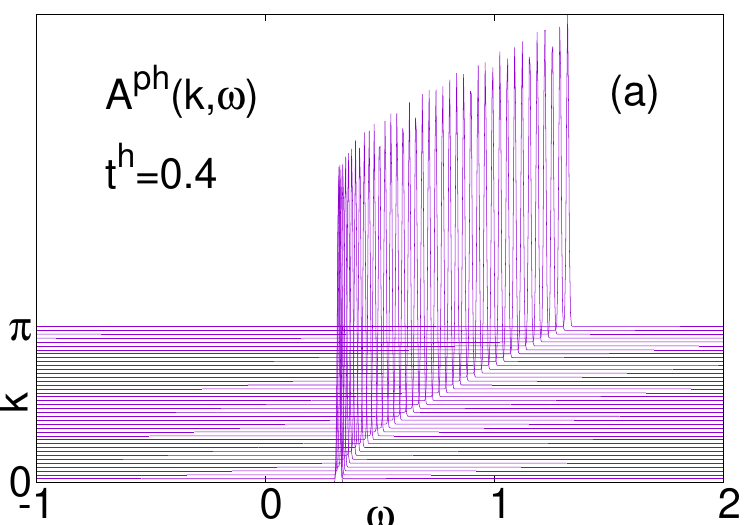}
\includegraphics[width=0.275\textwidth]{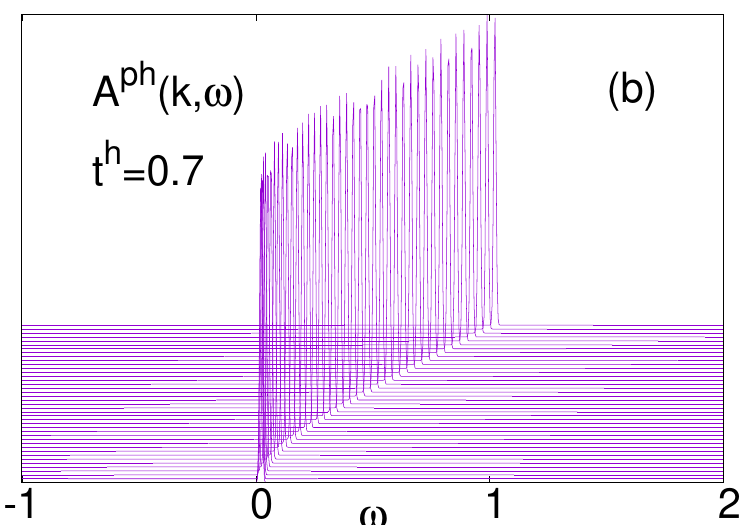}
\includegraphics[width=0.275\textwidth]{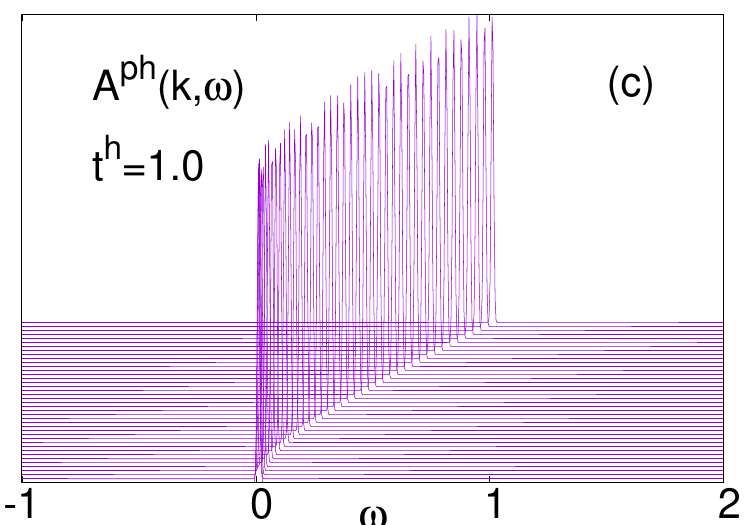}
\caption{Wave-number resolved photoemission spectra of photons $A^\textrm{ph}({\bf k},\omega)$ for the parameters set in Fig.~\ref{fig3}.}
\label{fig5}
\end{figure*}

Figure~\ref{fig4} shows us the intensity of the single-particle spectral functions of electrons, holes, and photons in the system along the $(k=k_x=k_y)$ direction in the first Brillouin zone for several values of $t^h$ at $U=1$, $d=2$, and $n=0.15$. Here the spectral functions are symmetry in momentum so we show in Fig.~\ref{fig4} only with $k\ge 0$. Once the mass of the holes is large ($t^h=0.2$ for instance), the \cor{valence} hole band is nearly flat [see Fig.~\ref{fig4}(b)] and one finds the more or less uniform distribution of the spectral weight of the single particles [cf. Fig.~\ref{fig4}(a\&b)]. In this case, even being mediated by both the matter-light coupling and Coulomb interaction, no hybridization between electrons and holes is established and the quasi-particle energies are almost not renormalized. No gap at all presents in the single particle spectral functions. Sufficiently lowering the mass imbalance by entering the condensation states, a gap feature opens at the Fermi level [see the left panels]. A double peak structure around the Fermi level indicates that there are two distinct energy states where electrons and holes can coexist. This is a hallmark of interactions leading to the hybridization or the bound states of the electrons in the conduction band and holes in the \cor{valence} band once the holes are more or less flexible. The width of the gap is proportional to $\Delta$ given in Eq.~\eqref{eq8} which is momentum independent. The double peak structure here suggests a robust coupling between the electrons and holes due to both the matter-light coupling and the electron-hole Coulomb attraction across the system. Increasing the hole mobility approaching that of electrons, the gap develops with respect to enlarging the excitonic and then the polaritonic coherences in the microcavity. 

In varying the mass imbalance of the electrons and holes in microcavity, we always find a single peak structure of the photon spectral function [see Fig.~\ref{fig5}]. Once the mass of holes is large, the localization tendency of the holes restricts the correlation between the electrons and holes, and no coherent states are established in the electron-hole-photon system. That induces the renormalized excitation photon band being far from the Fermi level [cf. Fig.~\ref{fig5}(a)]. However, by reducing the mass of the holes or increasing the holes' mobility to enter the excitonic condensation state, the matter-light coupling develops the exciton-photon coherence, specifying the roles of photons in stabilizing the coherent states in the system. Indeed, in the case of sufficiently large $t^h$, the photon spectral function shifts closely to the Fermi level, especially with momenta around ${\bf k}=0$ [cf. Fig~\ref{fig5}(b\&c)]. The development of the coherent electron-hole pairs with the photons by lowering the mass imbalance thus settles robustly the polariton condensation states in a microcavity.

\subsection{Susceptibility functions}

Inspecting the fluctuations of the coherent states is a significant point to understand their robust condensates. In the present work, we discuss here the static and dynamical excitonic and photonic susceptibility functions before the condensation states become stabilized. The static susceptibilities are delivered from their dynamical expressions [analytically represented in Eqs.~(\ref{eq17}) and (\ref{eq18}), respectively] in the limit of ${\omega\rightarrow 0}$ and then ${\bf q}\rightarrow 0$. Here we have assumed that in the directly semimetal/semiconducting situation, the bound states with respect to zero-momentum would be dominated~\cite{DP22}. \cor{Moreover, in general, the susceptibility at zero momentum corresponds to the uniform response, often termed the ``uniform susceptibility" which is critical for understanding global properties like magnetization, charge, or excitonic and photonic responses in our study under uniform fields. Note here that the susceptibilities in our work are considered in the uniform states, i.e., out of the order phases so it is reasonable to address the susceptibilities at zero momentum which is typical in studies of correlated electron systems~\cite{Georges06}. In RPA, the vertex corrections are all captured by ladder diagrams.}

\begin{figure}[hbt]
\includegraphics[width=0.490\textwidth]{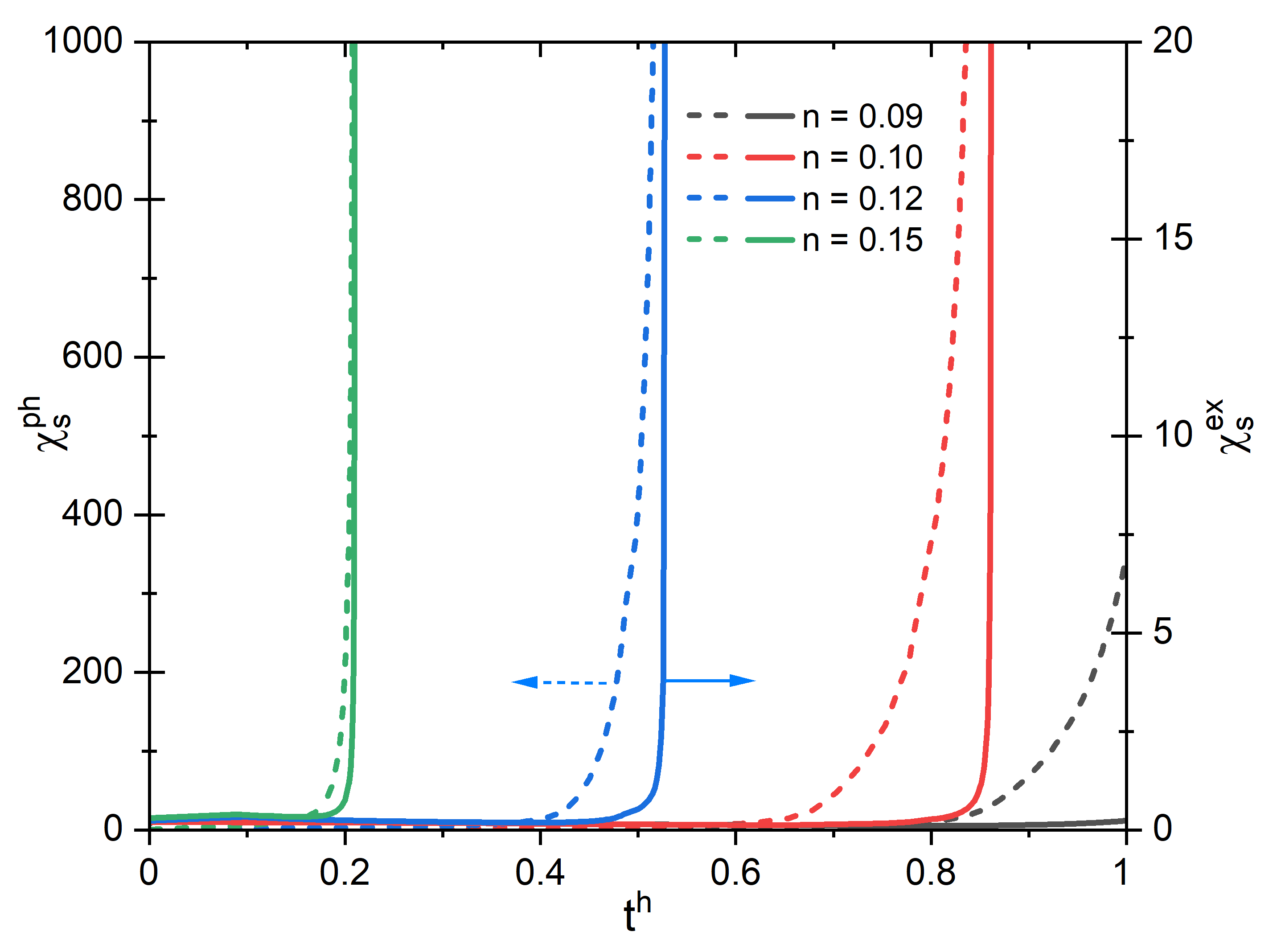}
\caption{Static excitonic ($\chi^\textrm{ex}_s$) and photonic ($\chi^\textrm{ph}_s$) susceptibility functions versus hole hopping integral $t^h$ for the sets of parameters in Fig.~\ref{fig1}.}
\label{fig6}
\end{figure}

Figure~\ref{fig6} shows us the static excitonic ($\chi^\textrm{ex}_s$) and static photonic ($\chi^\textrm{ph}_s$) susceptibility functions versus the hole hopping integral $t^h$ for Coulomb interaction $U=1$, detuning $d=2$, and at some different excitation densities as mentioned in Fig.~\ref{fig1}. Enlarging the hole hopping integral or lowering the mass imbalance reinforces the possibility of the electrons and holes coupling. For a given value of the excitation density, one thus always finds the development of both $\chi^\textrm{ex}_s$ and $\chi^\textrm{ph}_s$ as increasing $t^h$, indicating the fluctuated enhancement of the coherent bound states of both excitons and photons. Consequently, the quantum coherent polaritonic states tend to be robust once $t^h$ is sufficiently large. However, the tendency to the stability of the polariton condensates depends also on the excitation density. Indeed, at low excitation density, $n=0.09$ for instance, there is no divergence of the susceptibilities even in the limit of mass identity, indicating no robust polariton condensate is found (see black lines in Fig.~\ref{fig6}). Nevertheless, the situation is different such that the susceptibilities rapidly increase and then diverge at a critical value $t^h_c$ as increasing the hole hopping integral if the excitation density is larger. The divergence of the static susceptibilities specifies stability of the excitonic and photonic condensation states once $t^h>t^h_c$. Note here that, for a given value of the excitation density, both the excitonic and photonic susceptibilities diverge at the same position of the $t^h_c$. The instabilities of the excitonic and photonic condensates thus take place simultaneously, establishing the instability of the polariton condensation state. These signatures match very well with the results addressed before in Fig.~\ref{fig1}, at which both the excitonic and photonic order parameters concurrently become nonzero if $t^h>t^h_c$. Moreover, at the same hole hopping integral, one can also find that the photonic susceptibility is always dominant in comparison to that of the excitonic susceptibility. That scenario once more specifies the photonic fluctuations prominently over the excitonic fluctuations in the situation that the photonic level reaches the Fermi energy. This tendency specifies the major priority of the photonic fluctuations once the system reaches the polariton condensation state once the mass imbalance is small enough, especially at low excitation densities, as addressed in Fig.~\ref{fig1}.

\begin{figure}[hbt]
\includegraphics[width=0.480\textwidth]{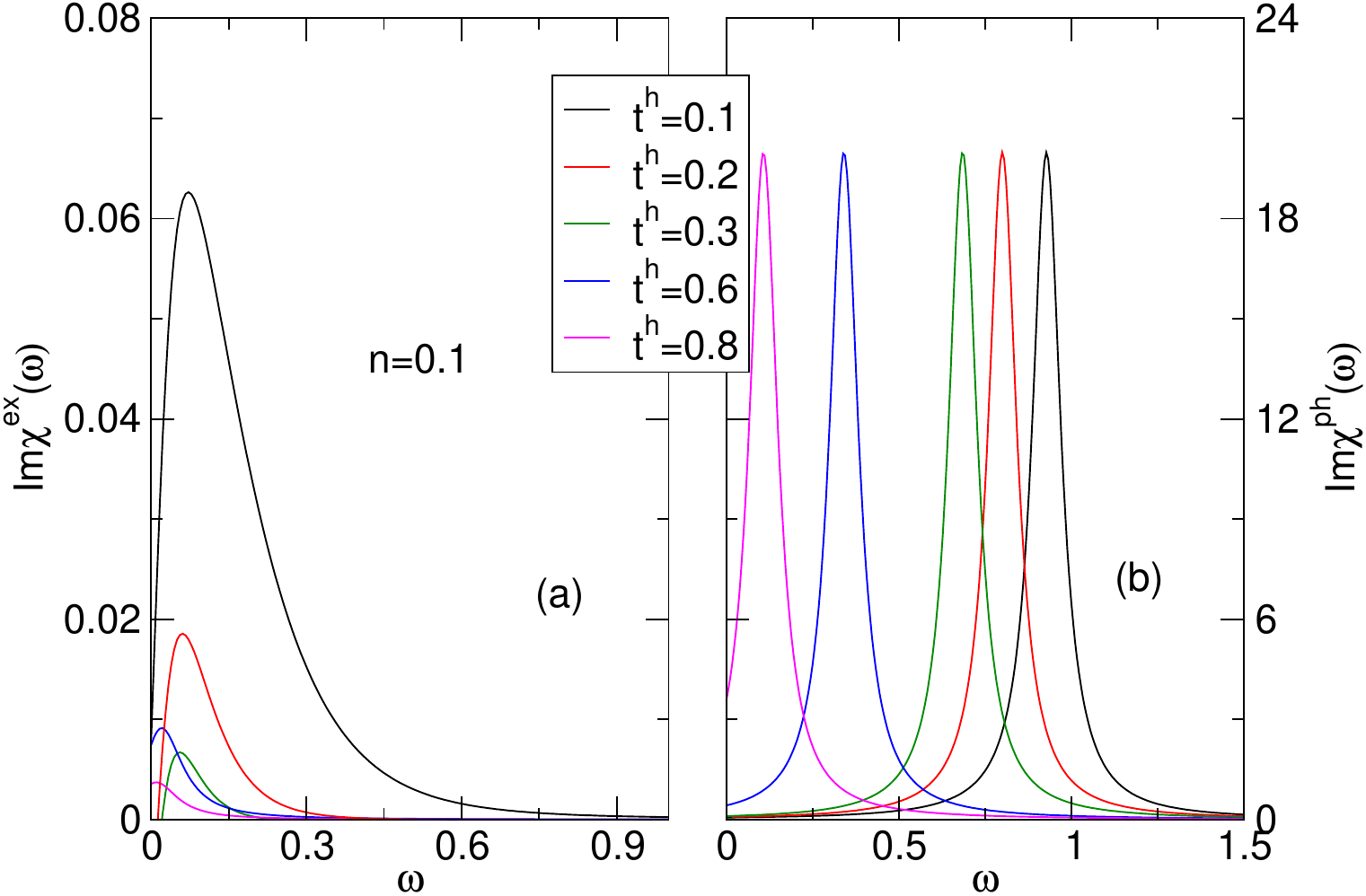}
\caption{The imaginary part of the dynamical excitonic [$\textrm{Im}\chi^{ex}(\omega)$--left panel] and photonic [$\textrm{Im}\chi^{ph}(\omega)$--right panel] susceptibility functions for some different values of hole hopping integral $t^h$ at excitation density $n=0.1$, Coulomb interaction $U=1$, and detuning $d=2$.}
\label{fig7}
\end{figure}

The properties of the coherent fluctuations in microcavity under the influence of mass imbalance can be analyzed more profoundly by inspecting the signatures of the dynamical excitonic and photonic susceptibility functions. Figure~\ref{fig7} addresses the imaginary part of the dynamical excitonic [$\textrm{Im}\chi^\textrm{ex}(\omega)$--left panel] and the dynamical photonic [$\textrm{Im}\chi^\textrm{ph}(\omega)$--right panel] susceptibility functions at zero momentum for Coulomb interaction $U=1$, detuning $d=2$ at several values of the hole hopping integral $t^h$ with specified excitation density $n=0.1$. In the set of parameters, we always find a single-peak structure in the imaginary part of the dynamical susceptibility spectra in the whole range of the mass imbalance. The single-peak signatures specify the coherent excitations in the system under the external perturbation. As increasing the hopping integral, both the peaks in the $\textrm{Im}\chi^\textrm{ex}(\omega)$ and $\textrm{Im}\chi^\textrm{ph}(\omega)$ shift to the left with lower frequency. The movement of the peak in the $\textrm{Im}\chi^\textrm{ex}(\omega)$ to the lower frequency indicates the lower energy cost required to excite holes in the \cor{valence} band coupling to the electrons in the conduction band to form excitons. The excitonic fluctuations thus are reinforced by lowering the mass imbalance. However, in the situation with larger mass imbalance, the holes like to be localized and thus the excitonic bound states might be stably formed, and one finds a higher peak in the $\textrm{Im}\chi^\textrm{ex}(\omega)$ spectrum. Lowering the mass imbalance also promotes the correlation between the electron-hole bound states with photons. The energy cost required for the exciton-photon coherent bound states thus decreases as lowering the mass imbalance. At the hopping integral $t^h\rightarrow t^h_c$, the $\textrm{Im}\chi^\textrm{ex}(\omega)$ and $\textrm{Im}\chi^\textrm{ph}(\omega)$ tend to peak at zero frequency, indicating the robust stability of the polariton condensation states. 

\begin{figure}[t]
\includegraphics[width=0.480\textwidth]{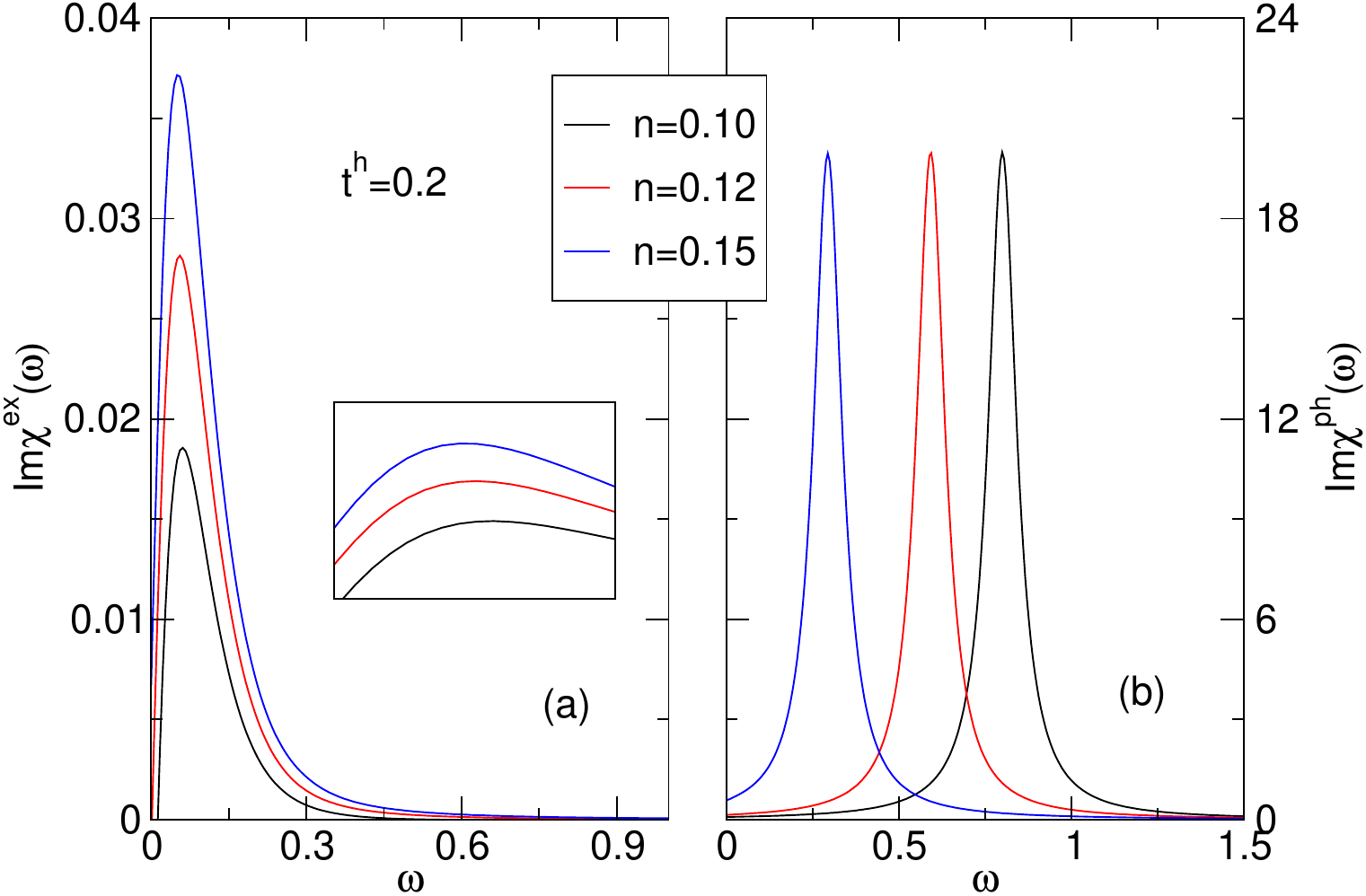}
\caption{The imaginary part of the dynamical excitonic [$\textrm{Im}\chi^{ex}(\omega)$--left panel] and photonic [$\textrm{Im}\chi^{ph}(\omega)$--right panel] susceptibility functions for some different values of excitation density $n$ at hole hopping integral $t^h=0.2$, Coulomb interaction $U=1$, and detuning $d=2$. The inset shows zoom in of $\textrm{Im}\chi^{ex}(\omega)$ close to their peaks.} 
\label{fig8}
\end{figure}

For a given mass imbalance situation, increasing the excitation density also releases the same feature of the bound-state fluctuations in the system. Indeed, Fig.~\ref{fig8} shows us that at the hopping integral $t^h=0.2$, both the $\textrm{Im}\chi^{ex}(\omega)$ and $\textrm{Im}\chi^{ph}(\omega)$ spectra display the peak-shift to the lower energy as increasing the excitation density in the range from $n=0.1$ to $n=0.15$. Increasing the excitation density, therefore, also stabilizes the excitonic coherent bound states, and thus promotes the polaritonic fluctuations in case of larger excitation densities.

\subsection{Phase diagrams}

\begin{figure}[t]
\includegraphics[width=0.490\textwidth]{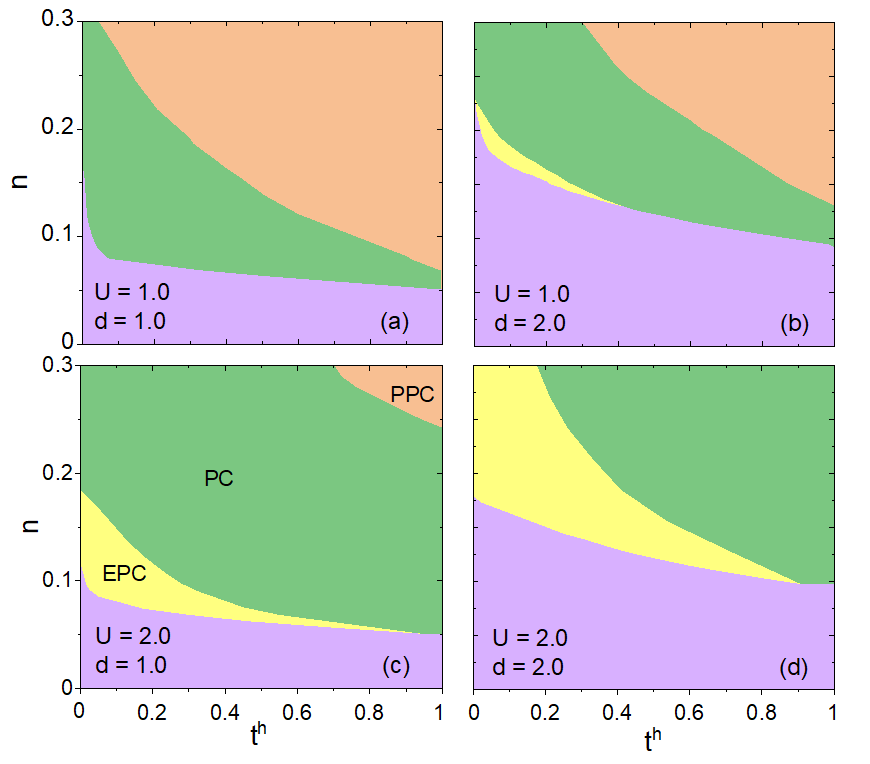}
\caption{Ground state phase diagram of the mass imbalance electron-hole-photon system in equilibrium microcavity in the $n-t^h$ plane for some different couple sets of Coulomb interaction $U$ and detuning $d$. The excitonic-like polariton condensate (EPC), the polariton condensate (PC), and photonic-like polariton condensate (PPC) are indicated respectively by the yellow, green, and orange regions. The purple indicates the normal disorder state of the electron-hole-photon system.}
\label{fig9}
\end{figure}

To analyze the stability of the condensation states in the mass imbalance microcavity, we summarize in Fig.~\ref{fig9} the ground state phase structure in the system in the $n-t^h$ plane for different set values of Coulomb interaction $U$ and detuning $d$. In Fig.~\ref{fig9}, the transition from the normal and disorder states to the condensates is specified by the divergence of the static susceptibility functions mentioned in Fig.~\ref{fig6}. Meanwhile, in order to examine in more detail the competition of the excitons and photons in establishing the polaritonic bound state we also address in the phase diagram the excitonic-like (EPC) and photonic-like (PPC) polariton condensates aside from the polariton condensation (PC) states. The PPC is specified if the ratio of the condensate order parameters: $\Delta_\textrm{ph}/\Delta>80\%$ and vice versa, the EPC is determined if $\Delta_\textrm{ph}/\Delta<20\%$, otherwise, the system settles in the PC~\cite{PRL.105.056401,Kamide2011}. 

With all couple sets of Coulomb interaction and detuning used in Fig.~\ref{fig9}, we always find the expansion of the condensates by lowering the mass imbalance. In the case of small Coulomb interaction $U=1$ and low detuning $d=1$, the PC is established suddenly if $t^h>t^h_c$ for a given sufficiently large excitation density $n$ [see Fig.~\ref{fig9}(a)]. The PPC is found once the excitation density is further increased. Increasing $t^h$ or lowering the mass imbalance, the development of hole mobility promotes the exciton-photon coherent states due to the matter-light coupling. The photons thus become the major part contributing to the polaritonic coherence. The PPC thus is expanded once the electron-hole system tends to the mass identity situation. Enlarging the detuning parameter has reinforced the overlap between the non-interacting hole \cor{valence} and electron conduction bands, that raises the electron-hole coherent bound state and one finds the EPC at low $t^h$ [see Fig.~\ref{fig9}(b)]. Increasing $t^h$ further, the contribution of photons to the polaritonic bound state becomes significant and the PC and then PPC are established by increasing the excitation density. At $U=2$ and $d=1$, one finds an expansion of the EPC especially at small $t^h$ or large mass imbalance [see Fig.~\ref{fig9}(c)]. In the situation, increasing the Coulomb interaction has promoted the electron-hole bound state and the excitons become significant once the hole \cor{valence} band is still in a narrow style. Lowering the mass imbalance or expanding the bandwidth of the hole \cor{valence} band, the photons gradually play their role in the polaritonic coherence in the presence of the matter-light coupling and the PC is established. Due to the larger Coulomb interaction, the PC is also expanded to the larger excitation density. The windows of the EPC and then the PC states are further extended in the case of larger detuning [cf. Fig.~\ref{fig9}(d)]. Note here that, enlarging the detuning has raised the overlap between the conduction and \cor{valence} bands, and as a consequency, the photons would play a less significant role in establishing the bound states at low excitation densities, that raises the critical values of the excitation density to stabilize the quantum coherent bound states in the microcavity. \cor{The result summarized in the phase diagrams fits quite well with the experimental observations at least in qualitatively. Indeed, at very low temperatures, studies on ZnO-based microcavities have specified the states of polaritonic condensation, characterized by an increase in the excitonic fraction (up to 96$\%$) and a decrease in the photonic fraction (down to 4$\%$) as the detuning is increased to 450 meV~\cite{PRL.110.196406}. This observation correlates well with our phase diagram at zero temperature, where the EPC becomes dominant with increasing detuning, while the PPC diminishes. This behavior aligns with experimental findings regarding the variations in photon and exciton fractions due to changes in detuning~\cite{Opt.2.1,PRB.107.165302,arXiv:2406.12940}. Moreover, the phase diagrams illustrated in Fig.~\ref{fig9} also highlight the evolution of PC and subsequently PPC states with increasing excitation density. This trend is consistent with experimental observations from GaAs quantum well microcavities, wherein an increase in excitation power leads to a transition from PC to a purely photonic laser state, characterized by a diminishing excitonic fraction~\cite{PRA.55.4632,PRL.121.047401}.}

\section{Conclusion}
To conclude, we have discussed the significant impact of the electron-hole mass imbalance on the polariton condensation states in a semimetal/semiconducting microcavity at ground state. The electron-hole-photon system is described in a two-dimensional thermal equilibrium many-particle model in which both the electron-hole Coulomb attraction and the matter-light coupling are taken into account on an equal footing. In the framework of the unrestricted Hartree--Fock approximation, we have derived a set of self-consistent equations evaluating the excitonic and photonic condensate order parameters. It specifies the mutual relation between the excitonic condensate order parameter and that of the photonic condensate. The polariton condensate thus appears as long as the excitonic or photonic condensate settles. With the help of the random phase approximation, both the exciton and photon dynamical susceptibility functions are analytically addressed. Our numerical solutions release that lowering the mass imbalance, the electron-hole bound state reaches the bottom of the photonic band and the polariton condensate becomes robust. The polariton condensate characterizes the excitonic condensation state once the mass imbalance is large, whereas, it typifies the photonic one in the range of small mass imbalance. The impact of the mass imbalance on the fluctuations of the excitonic and photonic coherent bound states is also addressed in the signatures of the excitonic and photonic susceptibility functions both in static and dynamical descriptions.  Notably, lowering the mass imbalance results in a shift of the imaginary part of the dynamical susceptibility functions to lower frequencies, highlighting the strengthening of bound coherent states before they become robust at sufficiently small mass imbalances. The phase diagrams illustrate the expansion of the polariton condensate as the mass imbalance between electrons and holes in the microcavity decreases. Furthermore, increasing the Coulomb interaction and the detuning parameter also promotes the development of quantum coherent states within the system. \cor{Our results match well at least in qualitatively with the experimental observations of the polaritonic stabilities in some mass imbalanced optical microcavities.} These findings encourage further investigation into the nature of polariton condensation states, particularly their topological properties and coherent states in nonequilibrium situations in our future research.
 
\section*{Acknowledgement}
This research is funded by Hanoi University of Mining and Geology, Vietnam, under Grant No. T25-19. 
%
\end{document}